\documentclass[12pt,onecolumn]{IEEEtran}

\usepackage{cite}

\usepackage{graphicx}
\usepackage{amsmath}
\usepackage{amssymb}
\usepackage{amsthm}
\usepackage{color}

\newcommand{\Real}{\mathbb R}

\def\x{{\mathbf x}}

\newcommand{\vect}[1]{{\boldsymbol{\mathbf{#1}}}} % vector
\newcommand{\mat}[1]{{\boldsymbol{\mathbf{#1}}}} % matrix
\newcommand{\y}{{\vect y}}

\newcommand{\dataset}{{\cal D}}

\newcommand{\GP}[0]{\mathcal{GP}} 

\newcommand{\Normal}[0]{\mathcal{N}} 
 
\newcommand{\prob}{p}
 
\newcommand{\cut}[1]{} % cut out a part of the text

\begin{document}

\title{Mapping Leaf Area Index with a Smartphone and Gaussian Processes}

\author{Manuel Campos-Taberner,
        Franciso Javier García-Haro, Álvaro Moreno, María Amparo Gilabert, Sergio Sánchez-Ruiz, Beatriz Martínez
        and Gustau Camps-Valls,~\IEEEmembership{Senior~Member,~IEEE}% <-this % stops a space
\thanks{{\bf Manuscript published in IEEE Geoscience and Remote Sensing Letters, vol. 12, no. 12, pp. 2501-2505, Dec. 2015, doi: 10.1109/LGRS.2015.2488682.}}
\thanks{M. Campos-Taberner, F.Javier García-Haro, A. Moreno, M. A. Gilabert, S. Sánchez-Ruiz, B. Martínez are with Dpt. of Earth Physics and Thermodynamics, Faculty of Physics, Universitat de Val\`encia, Dr. Moliner, 46100 Burjassot, Val\`encia (Spain). E-mail: manuel.campos@uv.es\newline
\indent G. Camps-Valls is with the Image Processing Laboratory (IPL), Universitat de Val\`encia, Catedr\'atico A. Escardino, 46980 Paterna, Val\`encia (Spain). E-mail: gcamps@uv.es
}
}

\maketitle

\begin{abstract}
Leaf area index (LAI) is a key biophysical parameter used to determine foliage cover and crop growth in environmental studies. Smartphones are nowadays ubiquitous sensor devices with high computational power, moderate cost, and high-quality sensors. A smartphone app, called PocketLAI, was recently presented and tested for acquiring ground LAI estimates. In this letter, we explore the use of state-of-the-art nonlinear Gaussian process regression (GPR) to derive spatially explicit LAI estimates over rice using ground data from PocketLAI and Landsat 8 imagery. GPR has gained popularity in recent years because of their solid Bayesian foundations that offers not only high accuracy but also confidence intervals for the retrievals. We show the first LAI maps obtained with ground data from a smartphone combined with advanced machine learning. This work compares LAI predictions and confidence intervals of the retrievals obtained with PocketLAI to those obtained with classical instruments, such as digital hemispheric photography (DHP) and LI-COR LAI-2000. This letter shows that all three instruments got comparable result but the PocketLAI is far cheaper. The proposed methodology hence opens a wide range of possible applications at moderate cost.
\end{abstract}


\section{Introduction}

\IEEEPARstart{T}{he} estimation of biophysical parameters from remote sensing data is a key issue for monitoring crop properties. Leaf area index (LAI) has been defined as the total one-sided leaf area in relation to the ground or the total foliage surface area per unit of horizontally projected ground surface area~\cite{CHEN92}. Two main approaches have been used to quantify LAI from the ground, either {\em direct} or {\em indirect}. Direct methods require an effort in collecting an optimal sample size and estimating plant density, which involves destructive harvest techniques~\cite{Breda01112003}. Indirect methods estimate LAI using optical instruments through the computation of the radiation transmitted or the canopy gap fraction~\cite{Welles01091996}. Sensors like LAI-2000 Plant Canopy Analyzers (LI-COR, Inc., Nebraska, USA) measure the gap fraction from five different angles simultaneously. Digital hemispherical photography (DHP) is another indirect technique for computing the gap fraction through cameras with hemispherical lenses (fish-eye) coupled. This method estimates LAI from measurements of the gap fraction, defined as the fraction of sky seen from below the canopy (upwards photography) or fraction of soil seen from above (downwards photography). Both Plant Canopy Analyzers and DHP are some of the most widely used classical optical instruments for indirect LAI estimation~\cite{Breda01112003,Welles01091996}. Indirect methods actually compute an {\em effective} leaf area index (LAI$_{eff}$). In this work, LAI estimates refer to effective LAI values. The difference between the actual LAI and the effective LAI may be quantified by the clumping index $\Omega$ through LAI$_{eff}$ = $\Omega \times$LAI~\cite{Chen_clumping}. The clumping index is almost always less than $1$, with the exemption of very regularly spaced leaf distributions.

An alternative sensor device to estimate LAI may be currently in our hands. Smartphones are becoming an accessible daily taken instrument for most of the population\footnote{In 2014, the number of global users of mobile phones surpassed that of desktop computers. Even in underdeveloped countries, the use of smartphones raises at a much faster rates. Beyond worldwide adoption of this technology, interaction with the smartphones also increases: The average smartphone user downloads 3 apps per month.}. The high adoption rate of smartphones in today's Society, along with the increase in computational power and sensing capabilities is being exploited in many fields of science and engineering. 
Actually, the use of smartphone components such as global position system (GPS), camera, accelerometer, and core processing power makes them suitable for a number of purposes, including methods for indirect LAI estimation. Smartphone capabilities are growing up day by day making them possible future measuring instruments.
Recently, a mobile application (PocketLAI) has been successfully introduced for LAI estimation~\cite{Confalonieri201367}, which was further tested on paddy rice against commercial instruments, such as LI-COR LAI-2000 and Decagon AccuPAR ceptometer, and during an entire rice crop season against digital hemispherical photography (DHP) and LAI-2000 in~\cite{Campos1502}.

Several methods such as physical, statistical, empirical, and hybrid methods have been used to deal with the bio-physical parameter estimation~\cite{Camps-Valls2011}. In this study, we will face the problem of LAI estimation following a modern statistical approximation. Statistical methods based on parametric approaches, such as vegetation indices, use explicit parametric equations and need prior physical knowledge. On the other hand, nonparametric methods do not need in principle any prior about the relationship between data, but they infer those relations directly from data analysis. In general, parametric methods for bio-physical parameters retrieval become less effective in terms of accuracy, bias and goodness-of-fit than nonparametric methods~\cite{Lazaro-Gredilla2014}.

In the framework of the nonparametric methods, Bayesian approaches have become an alternative proposal to other machine learning methods such as neural networks~\cite{Haykin1998} (NNs) or support vector machines~\cite{Scholkopf2001,campsvalls09semisvr}. Gaussian processes regression~\cite{Rasmussen2006} is been widely used for bio-physical parameters estimation in many remote sensing studies including chlorophyll content retrieval~\cite{Bazi2012}, solar irradiation~\cite{Salcedo-Sanz2014}, vegetation properties~\cite{Lazaro-Gredilla2014}. Besides of the robustness and stability, one of the characteristics that make GPR a particular useful tool is the combination of very good prediction accuracy and the ability to provide confidence intervals for the estimates. 

The remainder of the letter is organized as follows. Section 2 briefly describes the theory of the GPR. The methodology followed in this study is outlined in Section 3, describing both the ground dataset and the Landsat 8 images. Section 4 discusses the results obtained, and finally Section 5 concludes the letter with a discussion and outline of the future research.

\section{Gaussian Process regression (GPR)}

Standard regression approximates observations (often referred to as \emph{outputs}) $\{y_n\}_{n=1}^{N}$  as the sum of some unknown latent function $f(\x)$ of the inputs $\{\x_n \in\Real^D \}_{n=1}^{N}$ plus \emph{constant power} Gaussian noise, i.e. $y_n = f(\x_n) + \varepsilon_n,~\varepsilon_n \sim\Normal(0,\sigma^2)$. Instead of proposing a parametric form for $f(\x)$ and learning its parameters in order to fit observed data well, GP regression proceeds in a Bayesian, non-parametric way. A zero mean\footnote{It is customary to subtract the sample mean to data $\{y_n\}_{n=1}^N$, and then to assume a zero mean model.} GP prior is placed on the latent function $f(\vect{x})$ and a Gaussian prior is used for each latent noise term $\varepsilon_n$,
$f(\vect{x})\;\sim\;\GP(\vect{0}, k_\vect{\theta}(\vect{x},\vect{x}'))$,
where $k_\vect{\theta}(\vect{x},\vect{x}')$ is a covariance function parameterized by $\vect{\theta}$, and $\sigma^2$ is a hyperparameter that specifies the noise power.
Essentially, a Gaussian process is a stochastic process whose marginals are distributed as a multivariate Gaussian. In particular, given the priors $\GP$, samples drawn from $f(\x)$ at the set of locations $\{\x_n\}_{n=1}^N$ follow a joint multivariate Gaussian with zero mean and covariance matrix $\mat{K_\vect{ff}}$ with  $[\mat{K_\vect{ff}}]_{ij} = k_\vect{\theta}(\vect{x}_i,\vect{x}_j)$.

If we consider a test location $\x_*$ with corresponding output $y_*$, the $\GP$ defines a joint %the following  joint
prior distribution between the observations $\y \equiv \{y_n\}_{n=1}^N$ and $y_*$.
\iffalse
\begin{equation*}
  \left[\!\!
    \begin{array}{c}
      \vect{y} \\
      y_*
    \end{array}
    \!\!\right]
  \;\sim\;
  \Normal\left( \vect{0},\;\left[\!\!
      \begin{array}{cc}
        \mat{K}_{\vect{ff}}+\sigma^2\mat{I}_n & \vect{k}_{\vect{f}*}\\
        \vect{k}_{\vect{f}*}^\top & k_{**}+\sigma^2\\
      \end{array}
      \!\!\right]\right).
      \label{eq:jointprior}
\end{equation*}
\fi
Collecting available data in $\dataset\equiv\{\vect{x}_n,y_n|n=1,\ldots
N\}$, it is possible to analytically compute the posterior distribution over the unknown output $y_*$:
%
%\begin{subequations}
%\label{eq:preddist}
\begin{align*}
 %\label{eq:preddista}
 \prob(y_*|\vect{x}_*,\dataset)&=\Normal(y_*|\mu_{\text{GP}*},\sigma_{\text{GP}*}^2)\\
%\label{eq:preddistb}
\mu_{\text{GP}*} &= \vect{k}_{\vect{f}*}^\top (\mat{K}_{\vect{ff}}+\sigma^2\mat{I}_n)^{-1}\vect{
y} = \vect{k}_{\vect{f}*}^\top\boldsymbol{\alpha} \\
%\label{eq:preddistc}
\sigma_{\text{GP}*}^2 &= \sigma^2+k_{**}-
     \vect{k}_{\vect{f}*}^\top (\mat{K}_{\vect{ff}}+\sigma^2\mat{I}_n)^{-1}\vect{k}_{\vect
{f}*}.
\end{align*} 

GPs offer some advantages over other regression methods. Since they yield a full posterior predictive distribution over $y_*$, it is possible to obtain not only mean predictions for test data, $\mu_{\text{GP}*}$, but also the so-called ``error-bars'', $\sigma_{\text{GP}*}^2$, assessing the uncertainty of the mean prediction. The whole procedure only depends on a very small set of hyper-parameters, which combats overfitting efficiently. Also, inference of the hyper-parameters and the weights $\boldsymbol{\alpha}$ can be performed using continuous optimization of the evidence.
Note, however, that the bottleneck of the algorithm is the definition of the covariance (kernel, or Gram) function $k$: this function should capture the similarity between data instances. A standard, widely used covariance function is the {\em isotropic} squared exponential (SE), $k(\vect{x}_i,\vect{x}_j) = \exp(-\|\vect{x}_i-\vect{x}_j\|^2/(2\sigma^2))$, which captures sample similarity in most of the data problems efficiently. In this paper, we use the GPML MATLAB toolbox for the experiments\footnote{http://www.gaussianprocess.org/gpml/code/matlab/doc/}~\cite{Rasmussen2006}, which is also available along with many other regression methods in the simpleR MATLAB toolbox\footnote{http://www.uv.es/gcamps/code/simpleR.html}.

\section{Data collection and methodology}

This section covers the data collection of both field data and remote sensing images in the study, and reviews the adopted methodology for training and validating the GPR models.

\subsection{Field data}

GPR has been trained with the ground data acquired during the 2014 ERMES (An Earth Observation Model based information RicE Service) field campaign in Spain\footnote{http://www.ermes-fp7space.eu/}. LAI measurements were taken over selected farms within the rice district of Sueca (39°16’N, 0° 18’W) situated in the Albufera natural park (South of Valencia city, East of Spain). The area has a typical Mediterranean climate, mild, with an average annual humidity of 65\%. The average annual temperature is 17°C. Their mean values ranging from 11°C in January and 27°C in August. The mean annual precipitation is approximately 430 mm, which tends to be intense and concentrated in autumn.

The site is a homogeneous rice planting area of approximately 10 km $\times$ 20 km extension. Most of the paddy rice fields are rectangular and flat. The rice cropping practices are common on all the rice district. The field campaign was carried out on 26 Elementary Sampling Units (ESUs) from June the 17th to September the 8th during 10 days covering the entire rice season. Measurement dates were selected matching with Landsat 8 overpasses. ESUs were located at least 30 m away from the field borders and were approximately 20 m $\times$ 20 m size. The center of the ESU was geolocated using a GPS. Over each ESU, 16 photographs were acquired with the DHP and subsequently processed using the CAN-EYE software. In this study, the standard procedure for DHP data processing suggested in~\cite{Demarez2008} was followed. LI-COR LAI-2000 was also used to estimate LAI by making three replications of one reading above and eight below the canopy for each measurement and ESU. In addition, LAI was also acquired with PocketLAI. A representative LAI measurement acquired with the smart app was obtained averaging 18 single measurements over an ESU. The mobile application was installed on a Samsung Galaxy S4 GT-I9505, with a Quad-Core 1.9 GHz processor and 2 GB RAM. The PocketLAI uses the smartphone's camera for taking images with a resolution of 4128$\times$3096 pixels.

During the field campaign 5 more ESUs per day were identified as non-vetetated land covers such as bare soils, water bodies, and roads. This approach was done in order to represent LAI$\approx$0 ESUs in the training set to avoid possible wrong mean estimates as well as very low confidence values retrieved by the GPR~\cite{Verrelst2013157}.

\begin{figure}[!t]
\centering
\includegraphics[width=16cm]{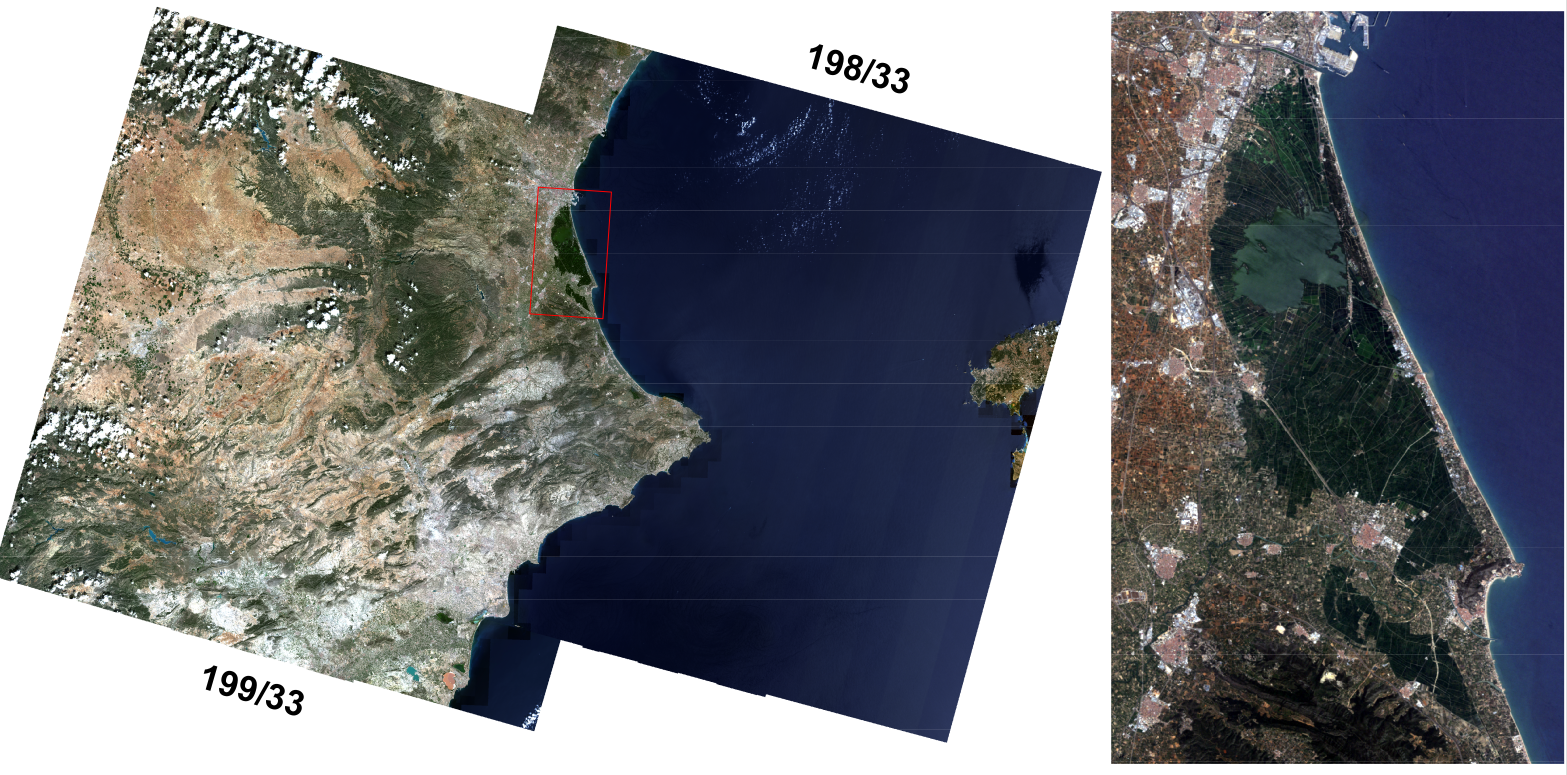}
\caption{RGB compositions of the two Landsat path/rows (198/33 and 199/33) (left), and the clipped image covering the rice area (right).}
\label{landsat8}
\end{figure}

\subsection{Landsat 8}

The Operational Land Imager (OLI) is a multispectral sensor on board Landsat 8 satellite. The spatial and temporal resolutions of Landsat 8 OLI images provide valuable information for crop monitoring at local scale~\cite{Campos15}. Recently, the United States Geological Survey (USGS) facilitated free access to Landsat archive data. In this work, Landsat 8 Surface Reflectance product was used for deriving high resolution LAI maps. Since the rice area lies in two Landsat path/rows (198/33 and 199/33), images were available every 7 and 9 days rather than the usual interval of 16 days. Six Landsat 8 OLI bands (B, G, R, NIR, SWIR1 and SWIR2) were selected to relate the surface reflectance with LAI measurements. Landsat 8 OLI images were clipped to 1500 $\times$ 800 pixel size covering the entire rice area (see Fig.~\ref{landsat8}). Pixels covering urban areas, sea and the lagoon were masked out during the retrieval process in order to avoid meaningless LAI estimates over those surfaces. 

\subsection{Training and testing}
During the entire rice growing season LAI was measured simultaneously using three instruments: PocketLAI, DHP and LI-COR LAI-2000. Hereafter we refer to each dataset as LAI$_{APP}$, LAI$_{DHP}$ and LAI$_{LIC}$ respectively. The three datasets and the associated Landsat 8 OLI surface reflectance were divided into two different training (80\%) and testing subsets (20\%). An independent model was built for LAI$_{APP}$,  LAI$_{DHP}$ and LAI$_{LIC}$. Each model was constructed by running GPR a hundred of times with different random selections of training and testing subsets. The testing subset was used for validation purposes evaluating the root-mean-squared error (RMSE) and the mean absolute error (MAE) to assess GPR accuracy. Mean error (ME) was used to evaluate the bias, and coefficient of determination ($R^{2}$) to account for the goodness-of-fit between predictions and measurements.

\section{Experimental results}

This section reports the experimental results of the study, both qualitatively through the generated explicit space-resolved high resolution maps, and quantitatively through measures of accuracy, fit and bias for the three LAI estimation tools.

\subsection{High resolution maps}

\begin{figure}[!t]
\centering
\includegraphics[width=16cm]{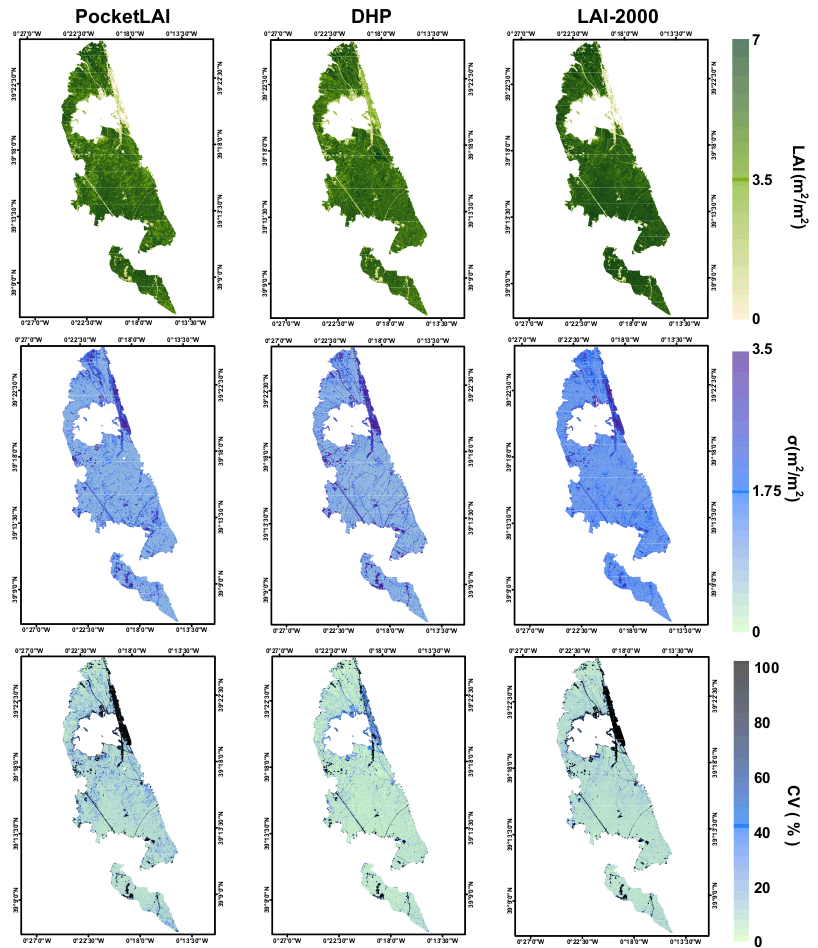}
\caption{(Top) Leaf area index prediction maps, (middle) standard deviation maps and (bottom), coefficient of variation, obtained with GPR using PcoketLAI, DHP and LAI-2000 ground data on the Landsat 8 image (DoY=212).}
\label{mapes}
\end{figure}

Landsat 8 OLI based LAI maps were derived for the rice district of Sueca (Val\`encia), Spain. Figure~\ref{mapes} shows the high resolution maps providing both mean estimate and associated uncertainties maps for \emph{LAI$_{APP}$}, \emph{LAI$_{DHP}$} and \emph{LAI$_{LIC}$}. We display generated maps on July 31th (DoY=212). Within-field variations are observed in all three LAI maps due the following reasons: (i) the spatial LAI variability of rice fields corresponding to different varieties, (ii) differences in plant phenological stages and (iii) low LAI values corresponding to non-vegetated areas and boundaries of the rice fields. In the rice fields, LAI estimates fell within the expected range at that phenological state. %~\cite{Fang2014126}.
Secondly, LAI uncertainty ($\sigma$) maps show low values within the rice fields. Higher uncertainties values appear over zones corresponding to non-vegetated areas or low LAI estimates. Particularly, on the east side of the maps it can be detected low LAI values and higher uncertainties referred to a different type of vegetation land cover (trees) and several man-made surfaces. This effect was also observed in~\cite{Verrelst2012}, which is essentially due to the fact that GPR cannot extrapolate outside the ranges seen in the dataset. Maps based on DHP ground data are able to predict more variability in LAI and uncertainty than the rest, and hence better identify rice field boundaries, roads, farm buildings, etc. Nevertheless, map based on PocketLAI is very similar in terms of LAI variability to LAI-2000. LAI-2000 provides uncertainties slightly higher but also higher values in terms of LAI. Uncertainty map values need to be interpreted as a confidence interval around the mean predictions. For this reason, we computed the coefficient of variation (ratio between LAI uncertainties and mean LAI predictions)
$CV = \sigma/\mu\ast100$ which provides relative uncertainties (see Fig.~\ref{mapes}). These maps show that the majority of rice pixels fall below the 20\% uncertainty threshold and can be considered well validated as proposed by the Global Climate Observing System (GCOS)~\cite{GCOS2011}.

\subsection{Statistical comparison}
\begin{figure}[!t]
\centering
\includegraphics[width=16cm]{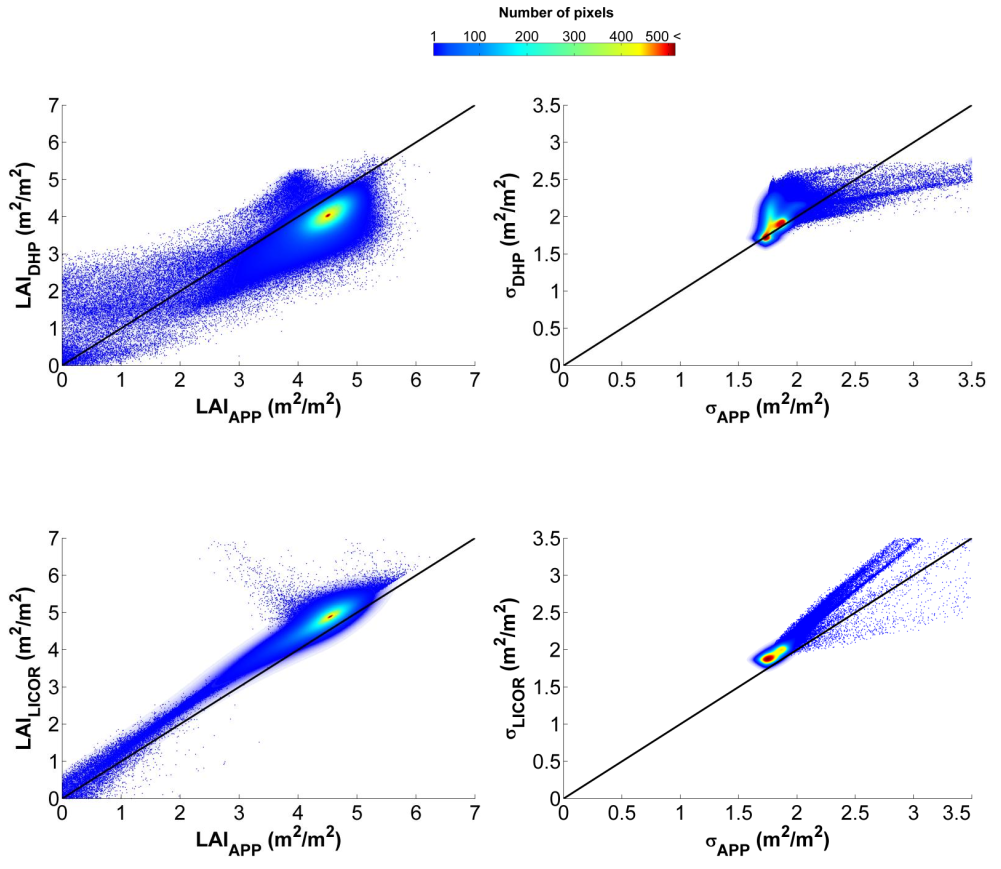}
\caption{Scatterplots of leaf area index mean estimates (left) and uncertainty (right) obtained with all three instruments.}
\label{scatter}
\end{figure}

Table 1 reports the statistical indicators of the models’ performance used for the LAI retrieval. GPR using LAI$_{APP}$, outperforms LAI$_{DHP}$ and LAI$_{LIC}$ models in terms of accuracy and goodness of fit. However, LAI$_{APP}$ bias is slightly higher than the other fits. The predictive mean estimates and associated predictive variances (uncertainties) between models were compared. Figure~\ref{scatter} shows the comparison of the scatterplots of the mean LAI estimates and associated uncertainties between models. In general, estimated LAI values with GPR from PocketLAI data were correlated to the classical instruments. Specifically, mean estimates retrieved when using PocketLAI showed similar results to those obtained with LAI-2000. Bias between LAI$_{APP}$-LAI$_{LICOR}$, LAI$_{APP}$-LAI$_{DHP}$, and LAI$_{DHP}$-LAI$_{LICOR}$ was 0.06, 0.05 and 0.12 in LAI units, respectively. This bias is slightly smaller than other reported in the literature for crops~\cite{Verger2009}. A comparison between associated uncertainties provided by the different models showed that the majority of pixels fall close to the 1:1 line, suggesting good fits. Therefore, using data acquired from PocketLAI do not produce high differences in uncertainties when compared with those produced by DHP and LAI-2000. The statistical confidence intervals of LAI predictions reliably identify areas that may be inaccurately mapped, such as man-made constructions and roads. Nevertheless, we would like to note that this is a conservative estimate of the LAI uncertainty since the error sources associated to experimental measurements and input reflectances are not well characterized. Modelling such error sources and designing proper GP priors are matters of active research, and will be pursued in future works.
\begin{table}[!t]
\small
\caption{Mean values of the statistical indicators (RMSE, MAE, absolute value of the ME, and R$^{2}$) between estimated and measured leaf area index in LAI$_{APP}$, LAI$_{DHP}$ and LAI$_{LIC}$ validation subsets. Standard deviations are shown in parenthesis.}
\label{table_example}
\centering
\begin{tabular}{|l|c|c|c|c|}
\hline
\hline
Dataset   &   RMSE &   MAE &  $|$ME$|$   &   R$^2$ \\
\hline
\hline
LAI$_{APP}$ & 0.51(0.18)&	0.35(0.19)&	0.12(0.07)&	0.94(0.07)\\
\hline
LAI$_{DHP}$ &	0.64(0.20)&	0.43(0.19)&	0.15(0.10)&	0.87(0.09)\\
\hline
LAI$_{LICOR}$  &	0.62(0.21)&	0.48(0.26)&	0.13(0.08)&	0.89(0.07)\\
\hline
\hline
\end{tabular}
\end{table}

\section{Discussion and future work}

This work introduced the use of a smartphone app, named PocketLAI, to estimate in-situ LAI. To assess the usefulness of the estimates, we then combined remote sensing Landsat 8 images with state-of-the-art non-parametric Gaussian process regression to generate spatially explicit high resolution LAI maps. We compared LAI predictions and confidence intervals for the estimates using PocketLAI and other classical instruments, such as DHP and LI-COR LAI-2000. The obtained results suggested that the proposed combination of a smartphone app and GPR was an appropriate cost-effective approach, in terms of accuracy, bias and goodness of fit.

We can conclude that, in general, the performance of the proposed approach based on PocketLAI smart app has shown to be less expensive in terms of instrument and processing. This avoids errors due to processing from non expert operators and makes it suitable for operational LAI monitoring activities. As we have shown, the confidence map obtained with LAI$_{APP}$ kept constant so thus the up-scaling process can be considered stable. These results allow considering PocketLAI as a powerful alternative to other commercial instruments not only for LAI monitoring during field campaigns but also for LAI retrieval. In any case, further studies are required on other crop types and using the app on other smartphones with different cameras, which could affect PocketLAI readings and hence LAI estimates.

PocketLAI may help in collecting a huge amount of data during the phenological cycle of vegetation, especialy over rice, thanks to the high portability and ease of handling. However, two shortcomings are observed here: the use of large datasets increases the computational effort needed for running GPRs, and the current GPR does not include any temporal information. Two methodological GPRs will be considered in the near future: sparse GPR models to cope with massive data, and temporal GPR to deal with non-stationarities, scales and trends of the acquired time series. 
Finally, we would like to mention that, to our knowledge, this is the first work combining smartphones and machine learning for biophysical parameter retrieval. The results are really encouraging and open a wide field for experimentation and biophysical parameter retrieval at affordable cost, both in time, computational and human resources. The forthcoming free of charge Sentinel-2 data  will be a perfect testbed for the proposed methodology. 

\section*{Acknowledgments}
The authors would like to thank the RESET CLIMATE 337 (CGL2012-35831), LSA SAF, and TIN2012-38102-C03-01 338 projects.


\begin{thebibliography}{10}
\providecommand{\url}[1]{#1}
\csname url@rmstyle\endcsname
\providecommand{\newblock}{\relax}
\providecommand{\bibinfo}[2]{#2}
\providecommand\BIBentrySTDinterwordspacing{\spaceskip=0pt\relax}
\providecommand\BIBentryALTinterwordstretchfactor{4}
\providecommand\BIBentryALTinterwordspacing{\spaceskip=\fontdimen2\font plus
\BIBentryALTinterwordstretchfactor\fontdimen3\font minus
  \fontdimen4\font\relax}
\providecommand\BIBforeignlanguage[2]{{%
\expandafter\ifx\csname l@#1\endcsname\relax
\typeout{** WARNING: IEEEtran.bst: No hyphenation pattern has been}%
\typeout{** loaded for the language `#1'. Using the pattern for}%
\typeout{** the default language instead.}%
\else
\language=\csname l@#1\endcsname
\fi
#2}}

\bibitem{CHEN92}
J.~M. Chen and T.~A. Black, ``Defining leaf area index for non-flat leaves,''
  \emph{Plant, Cell and Environment}, vol.~15, no.~4, pp. 421--429, 1992.

\bibitem{Breda01112003}
N.~J.~J. Breda, ``Ground-based measurements of leaf area index: a review of
  methods, instruments and current controversies,'' \emph{Journal of
  Experimental Botany}, vol.~54, no. 392, pp. 2403--2417, 2003.

\bibitem{Welles01091996}
J.~M. Welles and S.~Cohen, ``Canopy structure measurement by gap fraction
  analysis using commercial instrumentation,'' \emph{Journal of Experimental
  Botany}, vol.~47, no.~9, pp. 1335--1342, 1996.

\bibitem{Chen_clumping}
J.~Chen and T.~Black, ``Foliage area and architecture of plant canopies from
  sunfleck size distributions,'' \emph{Agricultural and Forest Meteorology},
  vol.~60, no. 3–4, pp. 249 -- 266, 1992.

\bibitem{Confalonieri201367}
R.~Confalonieri, M.~Foi, R.~Casa, S.~Aquaro, E.~Tona, M.~Peterle, A.~Boldini,
  G.~D. Carli, A.~Ferrari, G.~Finotto, T.~Guarneri, V.~Manzoni, E.~Movedi,
  A.~Nisoli, L.~Paleari, I.~Radici, M.~Suardi, D.~Veronesi, S.~Bregaglio,
  G.~Cappelli, M.~Chiodini, P.~Dominoni, C.~Francone, N.~Frasso, T.~Stella, and
  M.~Acutis, ``Development of an app for estimating leaf area index using a
  smartphone. trueness and precision determination and comparison with other
  indirect methods,'' \emph{Computers and Electronics in Agriculture}, vol.~96,
  no.~0, pp. 67 -- 74, 2013.

\bibitem{Campos1502}
M.~Campos-Taberner, F.~García-Haro, R.~Confalonieri, B.~Martínez, A.~Moreno,
  M.~Sánchez-Ruiz, S.and~Gilabert, F.~Camacho, M.~Boschetti, and L.~Busetto,
  ``Intercomparison of instruments for measuring leaf area index over rice,''
  in \emph{IEEE International Geoscience and Remote Sensing Symposium}, Milano,
  Italy, 2015.

\bibitem{Camps-Valls2011}
G.~Camps-Valls, D.~Tuia, L.~Gómez-Chova, S.~Jiménez, and J.~Malo, ``Remote
  sensing image processing,'' \emph{Synthesis Lectures on Image, Video, and
  Multimedia Processing}, vol.~5, no.~1, pp. 1--192, 2011.

\bibitem{Lazaro-Gredilla2014}
M.~Lazaro-Gredilla, M.~Titsias, J.~Verrelst, and G.~Camps-Valls, ``Retrieval of
  biophysical parameters with heteroscedastic gaussian processes,''
  \emph{Geoscience and Remote Sensing Letters, IEEE}, vol.~11, no.~4, pp.
  838--842, April 2014.

\bibitem{Haykin1998}
S.~Haykin, \emph{Neural Networks: A Comprehensive Foundation}, 2nd~ed.\hskip
  1em plus 0.5em minus 0.4em\relax Upper Saddle River, NJ, USA: Prentice Hall
  PTR, 1998.

\bibitem{Scholkopf2001}
B.~Scholkopf and A.~J. Smola, \emph{Learning with Kernels: Support Vector
  Machines, Regularization, Optimization, and Beyond}.\hskip 1em plus 0.5em
  minus 0.4em\relax Cambridge, MA, USA: MIT Press, 2001.

\bibitem{campsvalls09semisvr}
G.~Camps-Valls, J.~Mu\~noz Mar\'{\i}, L.~G\'omez-Chova, K.~Richter, and
  J.~Calpe-Maravilla, ``Biophysical parameter estimation with a semi-supervised
  support vector machine,'' \emph{IEEE Geosc. Rem. Sens. Lett.}, vol.~6, no.~2,
  pp. 248--252, Feb 2009.

\bibitem{Rasmussen2006}
C.~Rasmussen and C.~Williams, \emph{\BIBforeignlanguage{en}{Gaussian Processes
  for Machine Learning}}, ser. Adaptive Computation and Machine Learning.\hskip
  1em plus 0.5em minus 0.4em\relax Cambridge, MA, USA: MIT Press, Jan. 2006.

\bibitem{Bazi2012}
Y.~Bazi, N.~Alajlan, and F.~Melgani, ``Improved estimation of water chlorophyll
  concentration with semisupervised gaussian process regression,''
  \emph{Geoscience and Remote Sensing, IEEE Transactions on}, vol.~50, no.~7,
  pp. 2733--2743, July 2012.

\bibitem{Salcedo-Sanz2014}
S.~Salcedo-Sanz, C.~Casanova-Mateo, J.~Munoz-Mari, and G.~Camps-Valls,
  ``Prediction of daily global solar irradiation using temporal gaussian
  processes,'' \emph{Geoscience and Remote Sensing Letters, IEEE}, vol.~11,
  no.~11, pp. 1936--1940, Nov 2014.

\bibitem{Demarez2008}
V.~Demarez, S.~Duthoit, F.~Baret, M.~Weiss, and G.~Dedieu, ``Estimation of leaf
  area and clumping indexes of crops with hemispherical photographs,''
  \emph{Agricultural and Forest Meteorology}, vol. 148, no.~4, pp. 644 -- 655,
  2008.

\bibitem{Verrelst2013157}
J.~Verrelst, J.~P. Rivera, J.~Moreno, and G.~Camps-Valls, ``Gaussian processes
  uncertainty estimates in experimental sentinel-2 lai and leaf chlorophyll
  content retrieval,'' \emph{ISPRS Journal of Photogrammetry and Remote
  Sensing}, vol.~86, no.~0, pp. 157 -- 167, 2013.

\bibitem{Campos15}
M.~Campos-Taberner, F.~García-Haro, A.~Moreno, M.~Gilabert, B.~Martínez,
  S.~Sánchez-Ruiz, and G.~Camps-Valls, ``Development of an earth observation
  processing chain for crop bio-physical parameters at local scale,'' in
  \emph{IEEE International Geoscience and Remote Sensing Symposium}, Milano,
  Italy, 2015.

\bibitem{Verrelst2012}
J.~Verrelst, L.~Alonso, G.~Camps-Valls, J.~Delegido, and J.~Moreno, ``Retrieval
  of vegetation biophysical parameters using gaussian process techniques,''
  \emph{Geoscience and Remote Sensing, IEEE Transactions on}, vol.~50, no.~5,
  pp. 1832--1843, May 2012.

\bibitem{GCOS2011}
\BIBentryALTinterwordspacing
GCOS, ``Systematic observation requirements for satellite-based products for
  climate, 2011 update, supplemental details to the satellite-based component
  of the implementation plan for the global observing system for climate in
  support of the unfccc (2010 update, gcos-154),'' Tech. Rep., 2011. [Online].
  Available: \url{http://www.wmo.int/pages/prog/gcos/Publications/gcos-154.pdf}
\BIBentrySTDinterwordspacing

\bibitem{Verger2009}
A.~Verger, B.~Martínez, F.~C. de~Coca, and F.~J. García-Haro, ``Accuracy
  assessment of fraction of vegetation cover and leaf area index estimates from
  pragmatic methods in a cropland area,'' \emph{International Journal of Remote
  Sensing}, vol.~30, no.~10, pp. 2685--2704, 2009.

\end{thebibliography}
\end{document}